\documentclass[prl,twocolumn,reprint,showpacs,preprintnumbers,amsmath,amsthm,amssymb,amsfont]{revtex4-1}
\usepackage{times}
\usepackage{graphicx}
\usepackage{dcolumn}
\usepackage{bm}
\usepackage{color}
\usepackage{subfigure}
\usepackage[version=4]{mhchem}

\begin{document}
\title{Epitaxial strain control of hole-doping induced phases in a multiferroic Mott insulator Bi$_2$FeCrO$_6$}
\author
{Paresh C. Rout$^{(1)}$ and Varadharajan Srinivasan$^{(1,2)}$}
\affiliation{(1) Department of Physics, Indian Institute of Science Education
and Research Bhopal, Bhopal 462 066, India}
\affiliation{(2) Department of Chemistry, Indian Institute of Science Education
and Research Bhopal, Bhopal 462 066, India}
\begin{abstract}
Epitaxial strain has been shown to drive structural phase transitions along with novel functionalities in perovskite-based thin-films. 
Aliovalent doping at the $A$-site can drive an insulator-to-metal and magnetic transitions in perovskites 
along with a variety of interesting structural and electronic phenomena. Using first-principles calculations, we demonstrate here, how coupling epitaxial strain with 
$A$-site hole doping in a multiferroic double perovskite, \ce{Bi2FeCrO6}, could lead to mitigation of issues related to anti-site defects and lowered magnetisation in 
thin-films of the material. We also show that epitaxial strain can be used to manipulate the hole states created by doping to induce half-metal to insulator, antipolar
to polar, antiferromagnetic to ferromagnetic, orbital ordering and charge ordering transitions. We also predict the formation of a half-metallic polar phase with a large
magnetic moment which could be of immense fundamental and technological significance. 
\end{abstract}
\pacs{
77.55.Nv, 75.50.Gg, 75.50.-i, 75.10.-b, 75.25.-j, 75.30.-m, 75.80.+q
}

\maketitle

Perovskites (with molecular formula $ABO_3$) have emerged as promising candidates for multiferroicity, a phenomenon characterized by the coexistence of ferroic orders
such as ferroelectricity, ferromagnetism and/or ferroelasticity along with a coupling of at least two of these orders within a single
phase~\cite{Scott_Nature_2006, cheong, Ramesh_Spaldin, Khomskii_2008, Li}. Of particular interest are magneto-electric multiferroics, where the transition metal ($B$) 
sites contribute to the magnetic moment while  the ferroelectric polarisation is usually associated with polar distortions induced by the $A$-site cations. These oxides are typically 
antiferromagnetic (AFM) Mott insulators in their ground state, thereby possessing very low magnetisation~\cite{Ramesh_Spaldin,BFO}. By populating the $B$-sites with TM ions 
having dissimilar magnetic moments, i.e. creating the so-called double perovskites, the magnetisation can be increased through ferrimagnetism while retaining ferroelectricity and 
magneto-electric coupling. A prominent example of this approach is Bi$_2$FeCrO$_6$ (BFCO) which was predicted to be a multiferroic in bulk form with a ground-state magnetic 
moment of 2 $\mu_B$ per formula unit~\cite{baettig} corresponding to a G-type AFM spin ordering. 

In practice, multiferroic oxides are often grown as epitaxial thin-films on oxide substrates. Epitaxial strain has been shown to drive structural phase 
transitions in thin-films, stabilise magnetic states~\cite{meng2018strain}, induce polar-to-nonpolar transitions in layered oxides~\cite{james},
stabilise multiferroicity in thin-films~\cite{Lee,strain2,Millis_PRB_2016}, and significantly influence ferroic properties of thin-films~\cite{Schlom_2014}.  
Recently, we have shown that, expitaxially grown (001) oriented BFCO thin-films are unstable to anti-site defects and prefer 
a C-type AFM (C-AFM) ordered ground state~\cite{Paresh}, both features leading to loss of magnetisation~\cite{shabadi, khare2}. 
The anti-site defects occur basically due to the similarity of charge and ionic radii of Fe$^{3+}$ and Cr$^{3+}$  ions.

Aliovalent chemical doping (introducing electrons or holes) in complex oxides is
another practical route for controlling many properties such as structure, magnetism,
magnetic transition temperature, polarisation, transport properties,
etc~\cite{Hwang,Navarro,Hiroshi,Ibarra,zu2014,hole1}.
Particularly, doping in the Mott regime directly changes the valence
state of the transition metal ion (e.g. manganites) while doping in the extreme
of the charge transfer regime (e.g. cuprates) results in the formation of holes
on the oxygen lattice~\cite{Imada,Mitchell}. 

In this study, we explore aliovalent doping of BFCO thin-films in conjunction with
epitaxial strain as a plausible route to recover multiferroic properties and encode other interesting functionalities.  
We found that mixing in Sr$^{2+}$ ions at the Bi$^{3+}$ sites in BFCO, which is a Mott insulator, primarily resulted in changes in the valence states
of the TM ions~\cite{Rozenberg_SciRep_2013}. This in turn led to differing ionic sizes at the $B$-sites and reduced cation disorder, thereby increasing the magnetisation. 
Furthermore, changes in the oxidation state of Cr atom also resulted in charge disproportionation (CD) and half-metallicity. Epitaxial strain played the
role of stabilizing these phases, under different doping concentrations, as well driving half-metal to insulator (HMIT) and antipolar to polar (AP-P) transitions. 
The combined effects of strain and doping resulted in a half-metal ferroelectric phase in BFCO. While ferroelectric metals have recently brought in much 
excitement~\cite{Shi_LiOsO3_2013,puggioni2014,Rondinelli2016} in the field, half-metal ferroelectrics are yet to be explored. Apart from their interesting magnetic properties, 
we anticipate that  spin-selective screening effects could lead to stabilization of finite polarisation within the half-metallic ferroelectric state, paving the way for new generation 
of spintronics devices and multipurpose switches. The choice of Sr for doping was motivated by the following facts: (a) the average tolerance
factor calculated~\cite{tollerance} to be 0.96 (for 50\% doping) and 0.93 (for 25\% doping) ($<$ 1) which suggests that $R$3 (polar) structure would be more 
stable over centrosymmetric structures~\cite{Ravindran_PRB_2006, Xiang, Bellaiche_PRL_2015}, 
(b) \ce{SrCrO3} and \ce{SrFeO3} show many fascinating properties like orbital-ordering induced ferroelectricity~\cite{Kapil},
metal-insulator transition~\cite{Dang,Hee,Long}, and pressure-induced ferromagnetism~\cite{Toshiaki}. 

\begin{figure}[pbht!]
\includegraphics[clip=true,width=0.45\textwidth]{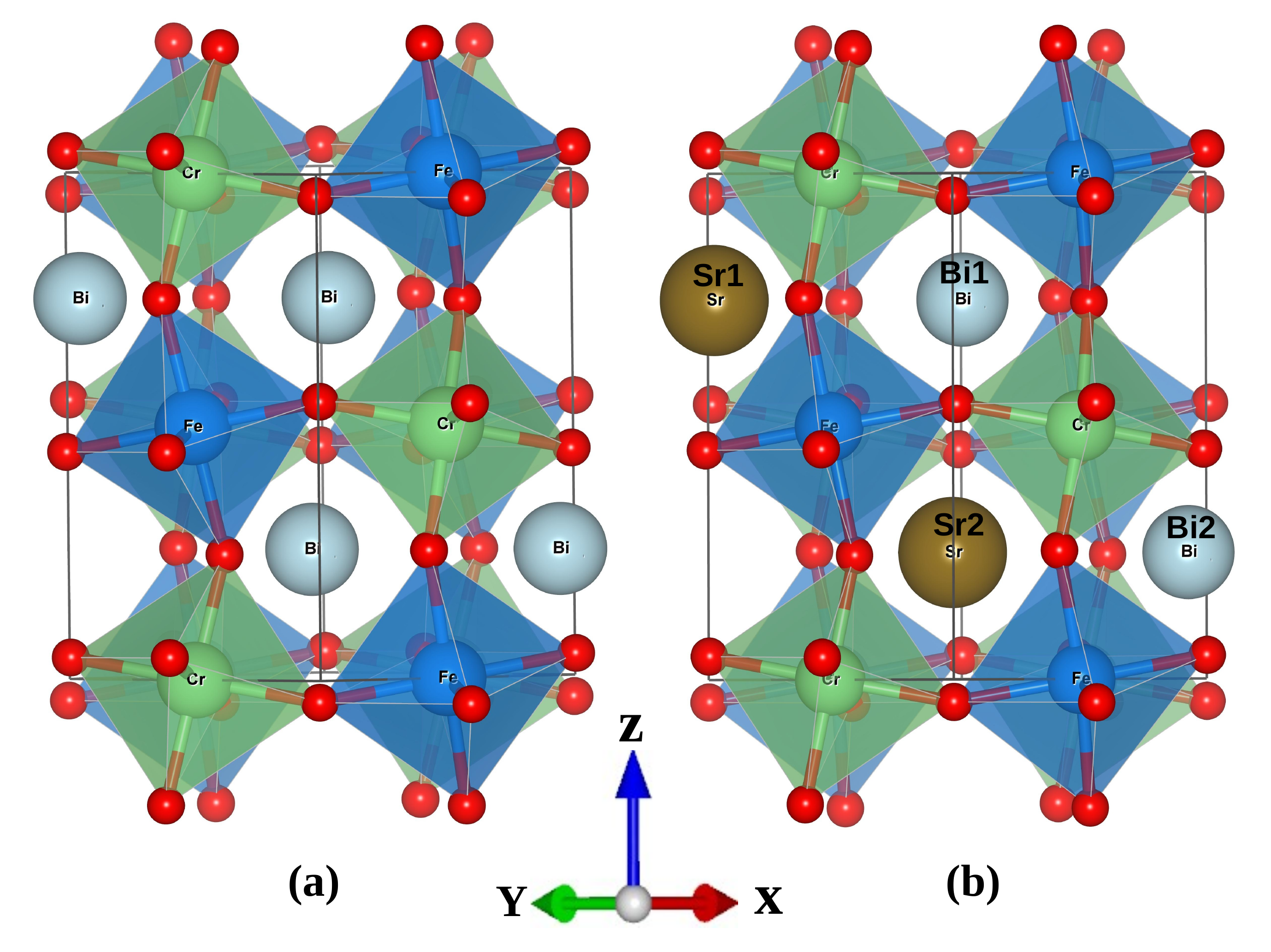}
\caption{The strained-bulk tetragonal supercell used in the calculations for modeling the (001) oriented
thin-films of (a) the parent \ce{Bi2FeCrO6} compound, and (b) of the 50\% doped compound, \ce{BiSrFeCrO6}, both with the
$a^-a^-c^+$ oxygen octahedra tilt pattern corresponding to the $P2_1/n$ (antipolar) space group.}
\label{fig:1}
\end{figure}

All calculations used the DFT+$U$ method. The details and justification of all
the parameters employed along with the supporting calculations used to arrive
at this model are presented in the Supplementary Material (SM). Our methodology
consisted of strained bulk calculations on a 20-atom 
supercell~\cite{Paresh} of BFCO (Fig.~\ref{fig:1}). The in-plane lattice parameters
of the supercell were constrained to various values spanning a strain window of $-4\% \leq \epsilon \leq 4\%$ ($\epsilon = 0$ refers to pseduocubic bulk BFCO). For each such strain,
the out-of-plane parameter, the monoclinic angle ($\beta$) and all internal coordinates were fully relaxed. Bi$_{2-x}$Sr$_x$FeCrO$_6$ (SBFCO) was modeled by replacing 
one ($x=0.5$) and two ($x=1$) Bi ions by Sr, respectively. Various $B$-site cation orderings as well as 
magnetic orderings were considered as starting points for relaxation calculations. Among the cation orderings possible within the 20-atom supercell, we focus below
on two energetically important ones: the rocksalt ordered structure (D0) and a structure with layers of Fe and Cr alternating in the (001) direction (D1).  
For the magnetic orderings, we considered three types of AFM (A, C, G-type) and the FM orders. Additionally, the structures
were chosen to conform to $P2_1/n$ (antipolar) and $R$3 (polar) space groups, respectively. These essentially differ by the sense and relative magnitudes
of oxygen octahedral rotations as indicated by the Glazer notations $a^- a^-
c^+$ and $a^- a^- a^-$, respectively. While other space groups are possible, 
we chose the above two as they were lowest in energy (see SM) and compatible
with cubic substrates~\cite{khare1, nache}. 

\begin{figure*}[pbht!]
{
\centering
\includegraphics[width=\linewidth]{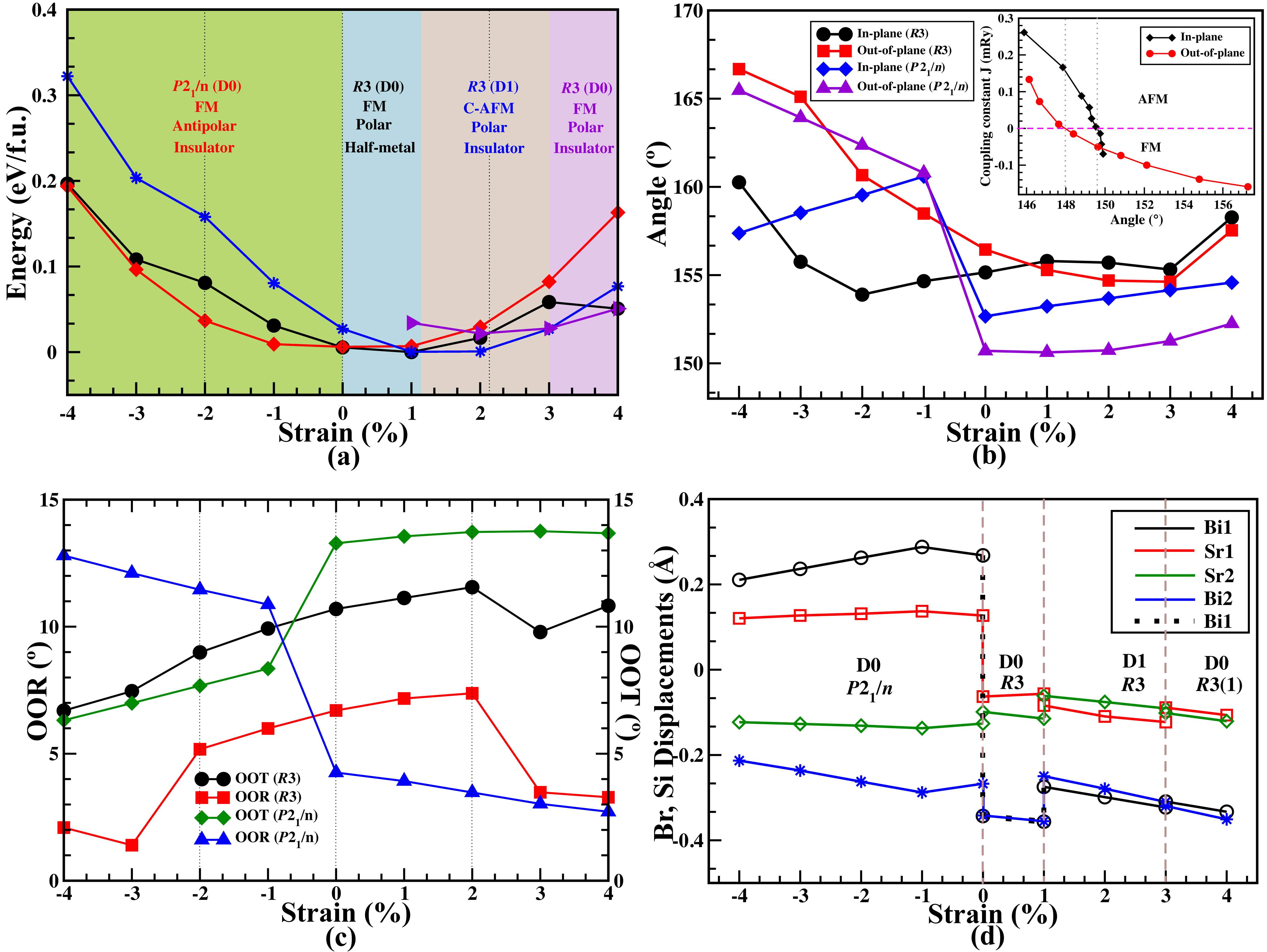}
}
\caption{Epitaxial strain induced phase transitions in BiSrFeCrO$_6$. (a) Total energy difference (in eV per formula unit) as a function of epitaxial
strain in the three lowest energy phases: black circles represent the FM state of D0 structure in $R$3 phase, red diamonds, the FM state of D0 structure in $P2_1/n$ -phase,
and blue asterisks, the C-AFM state of D1 structure in $R$3 phase. The indigo triangles represent the FM state of D0 structure in $R$3(1) phase which occurs as a result of an IPT at 2\% strain.
The dotted vertical lines indicate the locations of the isosymmetric phase transition (IPT) points (see part (c)); (b) In-plane (IP) and out-of-plane (OP) Fe-O-Cr angles of SBFCO structure: filled black circles
represent IP angle in D0 $R$3 phase, filled red squares OP angle in D0 $R$3 phase, filled blue diamonds, IP angle in $P2_1/n$ phase, and filled indigo triangles, OP angle
of $P2_1/n$ phase. The inset shows the variation of the IP and OP exchange coupling constant as a function of superexchange angle in epitaxially strained BFCO~\cite{Paresh};
(c) The average oxygen octahedral rotation (OOR) and oxygen octahedral tilt (OOT) angles as a function of biaxial strain for all the stable phases. The IPT (indicated by vertical
dotted lines) are characterised by sharp changes in the OOT and OOR in the antipolar ($P2_1/n$) phase at 0\% strain and in the polar ($R$3) phase at -2\% and +2\%
strain; and (d) The Bi, Sr displacements ( in \AA), along the pseudo-cubic [110] direction, from their respective positions in the ideal perovskite structure: open black circles
(as well as the dotted black line) represent Bi1 displacements, blue asterisks, Bi2 displacements, open red squares and green diamonds, Sr1 and Sr2 displacements,
respectively. The vertical dashed lines separate the various phases identified in part (a).}
\label{fig:2}
\end{figure*}

Our previous work~\cite{Paresh} (also see SM Fig.~22) found that undoped BFCO
thin-films tend to be susceptible to anti-site defects under all epitaxial
strains. Moreover, we also found that the lowest energy magnetic state for
coherently strained films was C-AFM ordered regardless of the cation ordering.
In doped BFCO we discovered that the situation changes drastically. We first obtained the optimized 
geometries in the $P2_1/n$ and $R$3 phases for different magnetic and cation
orderings as a function of epitaxial strain (see SM Figs. 2-4). Fig.~\ref{fig:2}(a) shows the energy 
(relative to the global minimum structure, i.e. $R3$ D0-FM at 1\% strain) 
of the lowest lying states as a function of epitaxial strain. The main observations in different strain regions 
are summarized below:
\begin{enumerate}
 \item \textbf{\underline{-4\% to 0\%}}: The antipolar ($P2_1/n$)
FM state (7$\mu_B$/f.u.) of D0 structure emerges as the stable state
of SBFCO structure. 
\item \textbf{\underline{0\% to 1.2\%}}: The polar ($R$3) phase of D0 structure
with FM ordering (7$\mu_B$/f.u.) is stable.
\item \textbf{\underline{1.2\% to 2.8\%}}: SBFCO is stable in the layered D1 
structure with the C-AFM order but with a polar ($R$3) structure.
\item \textbf{\underline{$>$2.8\%}}: The $R$3 FM phase of D0 structure once
again becomes the lowest energy phase.
\end{enumerate}

The immediate effect of hole-doping is the stabilisation of the rock-salt ordered D0 structure along with the realization
of a FM ground-state. Epitaxial strain further induces AP-P and FM to AFM transitions as discussed below. 
The sharp discontinuity in the energy of the D0 $R$3 phase around 3\% strain is indicative of a first-order phase transition. 
The two phases across this strain differed only  by a reorientation of octahedra maintaining the symmetry of D0 $R$3 
structure as a result of the constraints imposed by coherence and epitaxy. Such phase transitions are known as isosymmetric
phase transitions (IPT)~\cite{iso2,iso3,iso4} and are expected to be first order~\cite{iso1}. We refer to the new phase as
$R$3(1) below. The total energy curve of this FM $R$3(1) phase actually crosses the FM D0 $R$3 phase at $\approx$~2\%
strain, indicating that the IPT happens at this strain. A similar IPT is also observed at -2\% strain indicated by dashed vertical
in the D0 $R$3 phase structure. We also found an IPT at 0\% strain in the D0 $P2_1/n$ structure signalled by an abrupt 
increase in $c$/$a$ ratio of the cell (see SM Fig.~5). These IPT are also indicated by sudden changes in the average oxygen 
octahedral tilt (OOT) and rotation (OOR) angles~\cite{Zayak} in the D0 structure as shown in Fig.~\ref{fig:2}(c) for both 
$P2_1/n$ as well as $R$3 phases. The rotations and tilts are also accompanied by octahedral distortions which bring 
differentiation between the magnitudes of the various OOT and OOR angles within the 20-atom supercell (see SM Fig.~6). 
However, the abrupt changes at the IPT happen for all angles at the aforementioned strain values. A decomposition of the
structural changes into symmetry adapted normal modes revealed that all the IPTs occur along a single $\Gamma_1$ mode 
of the ideal structure (see SM Fig.~10).

The stabilisation of FM ordering in the D0 structure upon 50\% hole-doping is most directly an effect of the increase in the Fe-O-Cr super-exchange
angles compared to those in undoped BFCO. This is evident in Fig.~\ref{fig:2}(b) where the in-plane (IP) and out-of-plane (OP) Fe-O-Cr angles of 
the optimized D0 structures (for both $R$3 and $P2_1/n$ phase) are shown to be greater than 150$^\circ$ throughout all strains. For the Fe$^{3+}$ -
Cr$^{3+}$ ($d^5-d^3$) super-exchange system, the sign of the coupling constant switches at an angle of approximately 150$^\circ$ as shown 
previously~\cite{Paresh}. At angles beyond this critical value a FM exchange is obtained, while lower angles seem to yield AFM exchange. The
increase in these angles is related to the introduction of the larger Sr$^{2+}$ ions at the $A$-site, which leads to lesser tilts in the oxygen octahedra 
and, hence, larger Fe-O-Cr angles. Additionally, the hole doping results in an oxidation of the Cr site to a $d^2$ configuration further strenghtening
an FM super-exchange interaction~\cite{good,kanamori1,Wollan} between the Fe and Cr sites. Similarly, in the $R$3 D1 structure (1.2 - 2.8\% strain), 
the OP interactions are between Fe and Cr and become FM due to increased super-exchange angles ($\sim$157$^\circ$). However, the IP interactions
are between like ions (Cr-Cr and Fe-Fe) and are AFM in nature as expected from the Goodenough-Kanamori-Anderson rules (GKA)~\cite{good}. This
results in a C-AFM ground-state (AFM in the plane and FM out of the plane) in this region of strain. Hence, the FM to AFM transition at $\sim$1\% strain
is basically due to change in the $B$-site cation ordering. The abrupt changes seen in the super-exchange angles at 0\% strain in the $P2_1/n$ and at -2\%
strain in the $R$3 phase are associated with the IPTs mentioned above.  

In order to check the fate of the ferroelectric properties of BFCO upon hole-doping, we used the $A$-site ion displacement (relative to the ideal structure)
as an order parameter representing the polarisation~\cite{iso4}. Displacements of the Bi$^{3+}$ ions, driven by their stereochemically active lone pair,
contribute directly to the ferroelectric polarisation in BFCO~\cite{baettig}. The dependence of these displacements, computed along the pseudo-cubic
$\left[110\right]$ direction~\cite{Bellaiche_PRB_2013}, on the applied biaxial strain is depicted in Fig.~\ref{fig:2}(d) for the most stable phases
at each strain. A clear transition from antipolar to polar phase is seen at 0\% strain where the Bi and Sr displacements of SBFCO go from being anti-parallel
to each other to moving in the same direction. This is consistent with the change of space-group from $P2_1/n$ to $R3$ at 0\% strain. In fact, at 0\%
strain, starting a geometry relaxation from the $P2_1/n$ structure resulted in the $R$3 phase, indicating that the former phase is unstable at this strain.
Thus, an AP-P transition is induced by epitaxial strain upon hole-doping. The polar phase survives the cation re-ordering transition at 1.2\% thereby
dominating the entire tensile strain region.

\begin{figure}[pbht!]
\includegraphics[width=0.5\textwidth]{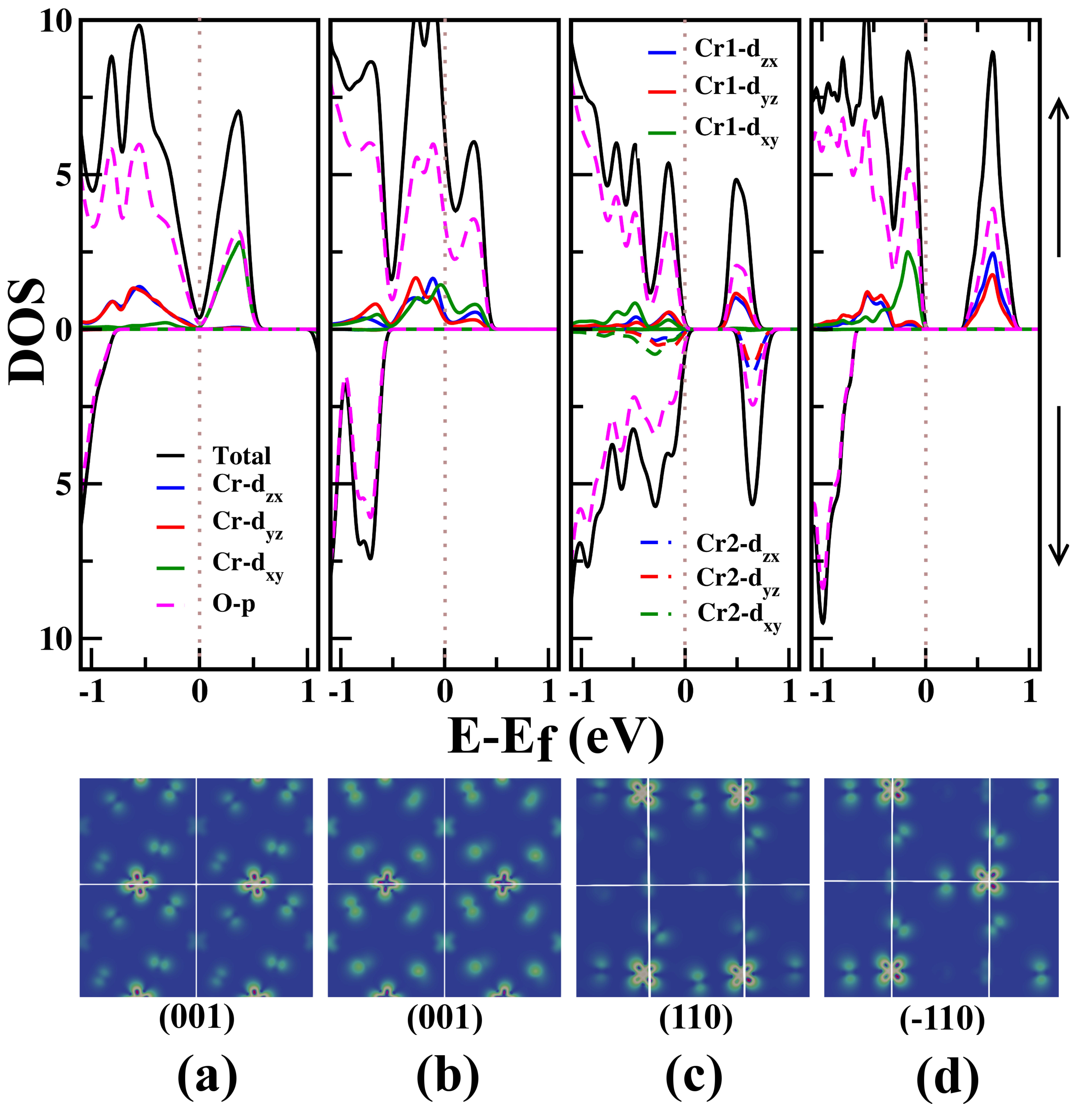}
\caption{Total and projected density of states (PDOS) of lowest energy phases of BiSrFeCrO$_6$ at various strains (top panel) and the spatial
distribution of the corresponding hole states (bottom panel) on selected planes (indicated below the figures). (a) The antipolar ($P2_1/n$) D0 phase at 
-4\% strain; (b) the polar ($R$3) D0 phase at 0\% strain, (c) the polar ($R$3) D1 phase at +2\% strain (Cr1 and Cr2 refer to the two spin inequivalent Cr
atoms in the system); and (d) the isosymmetric polar ($R$3(1)) D0 phase at +4\%.}
\label{fig:3}
\end{figure}

Fig.~\ref{fig:3} shows the density-of-states projected on the Cr $3d$ and O $2p$ atomic states (PDOS) which contribute the most near the Fermi energy 
(see SM Fig.~8 for PDOS in a wider range of energies). Hole-doping and epitaxial strain directly affect the Cr $t_{2g}$ orbitals which are strongly hybridised
with the O $2p$ orbitals. In the compressive strain region the $d_{xy}$ orbital is destabilised and splits off from the $t_{2g}$ triplet, appearing as the
narrow hole band in Fig.~\ref{fig:3}(a). The other two orbitals in the triplet remain nearly degenerate and are each singly occupied resulting in a
Cr$^{4+}$ ion with a moment of $\sim$~2$\mu_B$. Thus, the D0 FM $P2_1/n$ phase becomes insulating with a narrow gap of $\sim$~0.1 eV. 

With reducing compressive strain, the degeneracy of the three $t_{2g}$ levels is gradually restored, turning the system metallic and giving rise to an insulator 
to half-metal transition at 0\% strain. At the transition point, there is also a charge transfer from the O $2p$ to the Cr $3d$ giving the hole a dominant O $2p$
character (see Fig.~\ref{fig:3}~(b)). Thus, epitaxial strain pushes the hole-doped system to a negative charge transfer regime~\cite{Khomskii_Arxiv_2001,Zaanen_PRL_1985}.
Coincidentally, there is also a structural change from $P2_1/n$ to $R$3 phase at this strain. The HM $R$3 phase, which extends over a narrow strain region (0-1.2\%)
has an FM ordered ground state due to the same reasons as the $P2_1/n$ phase at compressive strain. The concurrence of half-metallicity and polar nature is,
however, a surprise. Sustaining polarisation in a metal has been considered unlikely because of screening by conduction electrons, making such materials
rare. While polar metals have been shown to exist~\cite{Rondinelli2016,puggioni2014}, there is no previous report of a polar half-metal. In such a material
spin-selective screening effects could stabilise an electrically polarised state, with dominant contribution to the polarisation arising from one-spin channel. 
This property would be extremely appealing for spintronics device applications and is currently under investigation. 

The cation-ordering transition seen at 1.2\% strain also coincides with a HMIT (see Fig.~\ref{fig:3}~(c)). Once again there is a clear splitting of the Cr
$t_{2g}$ levels, but this time the hole state corresponds to a linear combination of the $d_{xz}$ and $d_{yz}$ orbitals~\cite{CaCrO3, BaCrO3}. Another
linear combination of these orbitals is below the Fermi level along with the $d_{xy}$ level. The layered nature of the phase leads to an AFM interaction
between the Cr ions resulting in the C-AFM ordering. The hole state is localised on the Cr and nearby oxygens and is distributed mostly in a plane parallel
to (110). Interestingly, at larger strains ($>$2.8\%), where polar D0 phase re-emerges as the most stable phase $R$3(1),  the same disposition of the $d$-orbitals
survives. Hence, the HM nature of the polar D0 phase vanishes and we get an insulating ground state with a narrow band gap (see Fig.~\ref{fig:3}~(d)). 
The overall effect of strain is to generate orbitally ordered states as shown in the bottom panel of Fig.~\ref{fig:3} where the integrated local density of states
arising from the hole bands is plotted. From compressive to tensile strain the hole state continuously changes from being on the (001) plane to a perpendicular
plane. Thus, strain drives an orbital re-ordering transition that goes through the half-metallic state at 0\% with no orbital ordering (degenerate $t_{2g}$
levels). The IPT (originating at 2\%), driven by symmetry-preserving oxygen octahedral reorientations, leads to the HMIT in the polar D0 structure. 

\begin{figure*}[pbht!]
\includegraphics[clip=true,width=\textwidth]{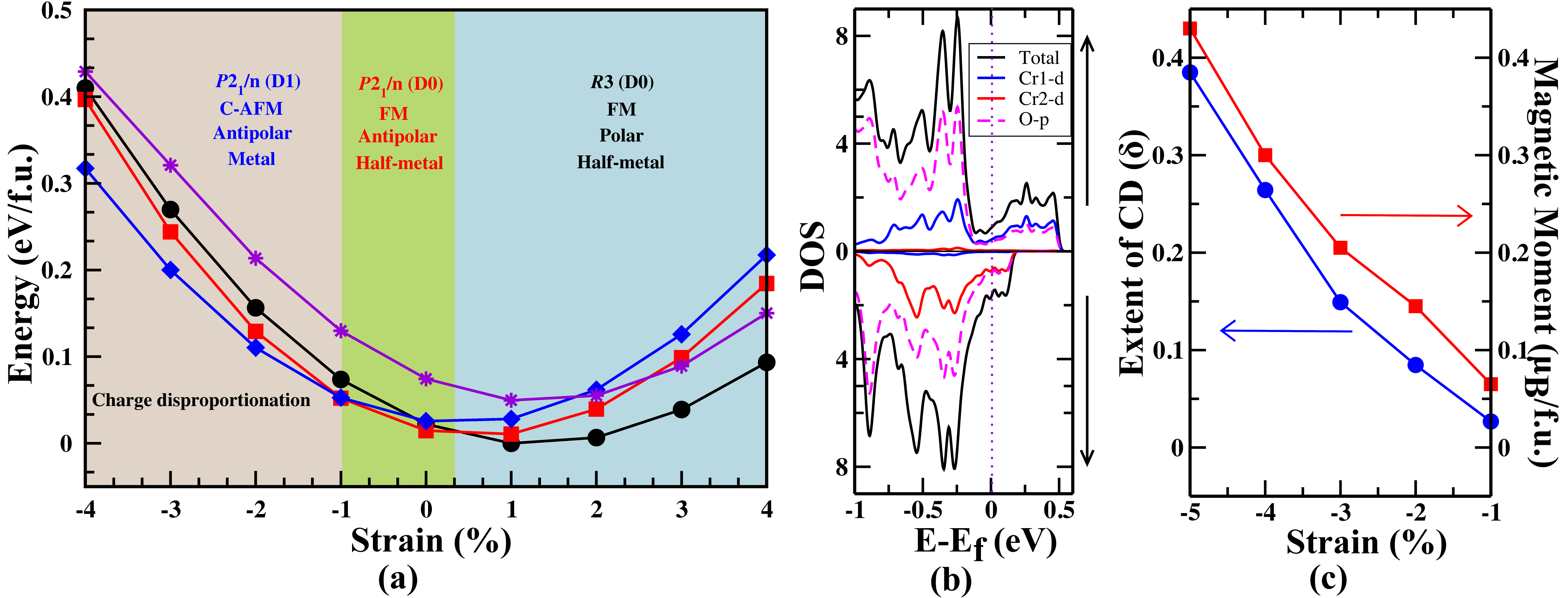}
\caption{Epitaxial strain induced phase transitions in Bi$_{1.5}$Sr$_{0.5}$FeCrO$_6$. (a) Evolution of total energy (in eV per formula unit) with epitaxial strain in the three lowest
energy phases: black circles represent the FM state of D0 $R$3 structure, red squares, the FM state of D0 structure $P2_1/n$ structure, and blue diamonds, the C-AFM state of D1
structure $P2_1/n$ structure. The energy of the C-AFM state of D1 structure in $R$3(1) phase, which was stable at 50\% doping, is also presented (indigo asterisks) for purposes of comparison.
All the energies are positioned with respect to the energy of the D0 FM (R$3$) state at 1\% strain, (b) Total and projected density of states (PDOS) of the antipolar
($P2_1/n$) phase of D1 structure at -4\% strain, and (c) The atomic charge difference between the two types of Cr atoms (Cr1 and Cr2) and magnetic moment per formula unit of C-AFM D1
structure as a function of compressive strain. }
\label{fig:4}
\end{figure*}

The polar HM phase, although interesting, is only stable in a narrow strain window. However, its stability can be enhanced by varying the stoichiometric 
percentage of Sr atom. A smaller Sr doping would also make charge disproportionation possible at the Cr sites  and could lead to charge-ordering
(CO). With this motivation in mind, we investigated the stability under epitaxial strain  of various phases of a 25\% doped BFCO structure (obtained by replacing
one-fourth of the Bi ions by Sr). The resulting phase diagram is summarised in Fig.~\ref{fig:4}. In the compressive strain region (i.e. -4\% to -1.0\% strain)
the C-AFM layered D1 structure becomes stable, like in the case of the pure BFCO, but is antipolar.  Between -1.0\% up to 0.4\% strain,
however, the antipolar FM phase becomes stable. Most surprisingly, the polar ($R$3) HM phase is stabilised in a wide range of tensile strain
(i.e. 0.4\% and beyond). The HM nature was confirmed by calculating its density of states (see SM Fig. 21). Thus, we not only found 
magnetic ordering transitions across strain but also observed structural transitions with 25\% Sr doping. However, the we did not observe any
strain-driven IPT in this case. Even with 25\% Sr doping the cation ordered D0 structure is stabilised under a wide range of epitaxial strain region
while retaining a large magnetization through a FM ordered ground state. 

Interestingly, we observed that, in the compressive strain region, the ground-state layered structure yields a small non-zero magnetic moment
($\sim$0.5$\mu_B$/f.u.) and is metallic despite an overall C-AFM ordering of the spins. Furthermore, this magnetic moment increases with 
compressive strain (see Fig.~\ref{fig:4}(c)) correlating well with an increase in the charge difference between the
two Cr sites in the layer. This is indicative of CD in the Cr sublattice to Cr$^{3.5+\delta}$ (Cr1) and Cr$^{3.5-\delta}$
(Cr2), respectively. At higher compressive strain the extent of CD ($\delta$) increases with the ions tending towards $4+$ ($t_{2g}^2 e_g^0$)
and ${3+}$ ($t_{2g}^3 e_g^0$) oxidation states, respectively. The CD is also confirmed by the PDOS (see Fig.~\ref{fig:4}~(b)) for the system
where the two Cr atoms are seen to be clearly inequivalent, not only having opposite spins but also contributing differently at the Fermi level. 
Furthermore, with increasing compressive strain the octahedral volume at Cr1 decreases much faster than that at Cr2 clearly indicating that
the former has a larger charge (see SM Figs. 19 and 20). The measured magnetic moment of $0.43~\mu_B$/f.u. at -4\% strain is consistent
with the proposed CD and the fact that the Fe ions are unaffected (Fe$^{3+}$:$t_{2g}^3$ $e_g^2$) by the doping. The localization of the hole
on the Cr layer leads to a checker-board patterned CO in the 20-atom supercell considered. 

Both the antipolar ($P2_1/n$) D0 phase and the polar ($R$3) D0 phase are HM as indicated by their PDOS (see SM Fig. 21).
Both phases have FM ordering with a total spin magnetic moment of 7.5 $\mu_B$/f.u. The Fe atoms retain their formal 
oxidation state of 3+ ($t_{2g}^3$ $e_g^2$) whereas the Cr atoms are equally divided between 4+ and 3+ oxidation states, with the hole localised
on the Cr atom closest to the Sr atom. This is the highest magnetic moment predicted in Bi-based double perovskites so far.

In summary, our calculations show that Sr doping of BFCO thin-films is a viable strategy to
remove cation disordering at the $B$-site. The size and charge modulation brought in by Sr not only ensures cation ordering
but also a FM ground-state with a large (7.5 $\mu_B$/f.u.) moment. At higher doping concentrations we found that epitaxial strain
could drive antipolar to polar, rock-salt to layered $B$-site ordering as well as isosymmetric phase transitions. These transitions coincide 
with AFM to FM, HM to insulator, and orbital ordering transitions. The most interesting observation was the stabilization of a 
half-metallic polar state in a narrow tensile strain window. The stability range of this phase is significantly enhanced when the doping concentration
was lowered to 25\%. At low hole doping, compressive strains stabilise the layered D1 structure with an overall C-AFM arrangement of spins.
The hole localises on the Cr layer inducing CD and, thereby, a charge-ordered state with a magnetic moment. The extent of CD and the moment were found 
to be proportional to the compressive strain. The CD renders the system metallic despite the AFM order. A cation ordering transition to the D0 structure occured at -1\% strain,
and a AP-P transition occurs past 0\% strain to result in the HM polar structure which remains stable across higher strains. Although our main aim behind
the hole-doping was to recover the ferroic properties, the interplay of the doping and epitaxial strain also makes BFCO host other interesting features. 
While the Fe ions retain their oxidation and spin states, the Cr ions accommodate the added holes and are ultimately responsible for the various electronic
transitions seen. Higher oxidation state in Cr is expected to result in interesting electronic behavior as a result of competitions between correlation effects and structural distortions~\cite{CaCrO3,SrCrO3,BaCrO3}.
Interestingly, the epitaxially strained Sr-doped film is able to host almost all the anticipated orders connected to each other by strain-induced phase transitions. 
The HM polar phase we have predicted has so far not been reported previously and is certain to attract much attention in the near future. We anticipate that the SBFCO structure 
can be readily formed in experiments as indicated by favourable tolerance factors as well as the fact that thin-films of the parent compound have already been formed. We hope that our 
observations will initiate further experimental efforts to verify the interesting predictions made in this work.

\acknowledgements{The authors would like to thank Mr. Shashi Saurav and all
the AITG group members at IISER, Bhopal for their help and invaluable
discussion. The authors gratefully acknowledge IISER Bhopal for computational
resources and funding. PCR would like to acknowledge CSIR for funding through
the JRF programme.}
\bibliographystyle{apsrev4-1}
\bibliography{./reference}
\end{document}